# Blockchain applications in Healthcare: A model for research

Amir Hussain Zolfaghari[1], Herbert Daly[2], Mahdi Nasiri[3], Roxana Sharifian[4*]

1. MSc in Medical Informatics, School of Management and Medical Information Sciences, Shiraz University of Medical Sciences, Shiraz, Iran.
2. Senior Lecturer in Computer Science at University of Wolverhampton, UK.
3. Assistant Professor in Artificial Intelligence, School of Management and Medical Information Sciences, Shiraz University of Medical Sciences, Shiraz, Iran.
4. Associate Professor in Health Information Management, Health Human Resource Research Center, Shiraz University of Medical Sciences, Shiraz, Iran: *Corresponding Author (Sharifianr@sums.ac.ir)

## Abstract

Blockchain technology has rapidly evolved from an enabling technology for cryptocurrencies to a potential solution to a wider range of problems found in data-centric and distributed systems. Interest in this area has encouraged many recent innovations to address challenges that traditional approaches of design have been unable to meet. Healthcare Information Systems with issues around privacy, interoperability, data integrity, and access control is potentially an area where blockchain technology may have a significant impact. Blockchain, however, is a meta-technology, combining multiple techniques, as it is often important to determine how best to separate concerns in the design and implementation of such systems. This paper proposes a layered approach for the organization of blockchain in healthcare applications. Key issues driving the adoption of this technology are explored. A model presenting the points in each layer is explored. Finally, we present an example of how the perspective we describe can improve the development of Health Information Systems.

## Keywords

Blockchain, Healthcare, Interoperability, Data Sharing, Software architecture infrastructure, Decentralization, Privacy

# Introduction

When people felt that the middleman should be disposed of the financial transactions, Bitcoin (1) was born via blockchain (2–4) technology. Blockchain was a technology that was not new, but brought with it available techniques to shape something new while the many attempts before (4) Bitcoin to create digital currencies could not draw people's attention. However, Bitcoin came with a secret of satisfying security and tamper-proof-ness in transactions with the help of modern cryptography, and that was how the first cryptocurrency was born.

After Bitcoin opened new doors for everyone to see blockchain technology capabilities, two happenings occurred in the following years. First, many startups attempted to introduce new cryptocurrencies by adding novel features or satisfying a particular need (5). Then, other industries found Blockchain, as the underlying technology of Bitcoin, to be right for solving issues that they are engaged in it.

One of the growing interests on the blockchain is in healthcare, where many experts (6–11) believe that decentralized technology could solve some of its issues. Thus, scientists found medical data availability beyond their expectations by applying blockchain in healthcare.

Patients' medical histories significantly influence the exact diagnosis (12–16), while many studies have proven electronic health records to be effective in improving the quality of treatment (17–21). Patients' locations might continuously change throughout their lives, and as a result, their records might get lost as they are saved in isolated databases [22]. Hence, the electronic health systems should be designed as interoperable with the ability to have continuous access throughout the patients' lifespan, with accessibility by health centers to medical histories.

However, before utilizing blockchain, we have to explore the current status of our software needs to see if the Decentralized Ledger Technology (DLT) can solve our issues. Some have suggested ways (22–25) to study the appropriateness of blockchain in each case before implementation. The data in healthcare is already segregated (26,27), as we have various information systems (28), diverse stakeholders (29), and different settings (28), consequently, integrating those entities by a DLT solution is logical. Moreover, the need to exchange a patient's history in various formats amongst different stakeholders requires a trustable and transparent system such as blockchain.

The rest of this study is organized as follows. First, some previous healthcare blockchain papers are briefly introduced in the "Related Studies" section. Next, we discuss why it is essential to regulate and standardize blockchain to improve the academic voice. Then, the traits of blockchain are outlined in seven layers. Finally, a "Functional Synthesis" of findings and study "Conclusion" comes afterward, plus suggestions for future works.

## Related Studies

Recently the interest in using blockchain and distributed ledger technology in healthcare is on the rise. Some published white papers provide a new model for health IT systems, and some others preferred scientific articles.

In a white paper published by Deloitte, Krawiec et al. (30) discuss that blockchain can empower interoperability as they consider the current healthcare records to be disjointed due to developmental and standard problems. While they presented some of the implementation challenges, it also stated that blockchain could be the leverage between healthcare stockholders.

Nichol et al. (31) showed the potential of blockchain technology in solving healthcare interoperability issues. They mention trust established by blockchain ecosystems can impact patient satisfaction positively as well as better healthcare outcomes by reducing the security risks.

A team at Vanderbilt University (32) recently proposed a healthcare blockchain model and named it FHIRChain. They attempt to satisfy "interoperability roadmap requirements" (33) regulated by ONC (The Office of the National Coordinator for Health Information Technology (ONC), a governmental organization in the US). In this study, Zhang et al. summarized essential technical prerequisites and expressed the proper blockchain solution for each requirement. They try to introduce a standard, secure, and scalable architecture, integrating with the current systems to provide data sharing for better health decision making. In the mentioned study, five data sharing issues indicated: patient identity, secure data exchange, access permission, data format, and software modularity. Then, the solutions were presented by the robust characteristics of blockchain technology. One outstanding feature in the FHIRChain model is its compatibility with HL7 FHIR (34,35) (Health Level Seven International health-care standards organization, Fast Healthcare Interoperability Resources), an international standard for electronic health record data exchange.

Unlike other studies, Zhang and Lin (36) designed their model to store data inside Blockchain. They hoped to use blockchain for secure and privacy-preserving PHI (Personal Health Information) while believing a better diagnosis would occur by sharing data, using blockchain technology. Their solution was to use two parallel blockchain networks; a "private blockchain" one for storing essential encrypted PHI data, and the other one a "consortium blockchain" to record data references.

Zhou et al. (37) identified that they stored data inside the blockchain, although many believe the in-chain data storage cannot lead to a comprehensive solution as long as we are dealing with multi-type data in healthcare. Some file types like X-Ray images are very hard to store that way. On the other hand, they mainly designed their model for the insurance data exchange plans since their system deals with light database records, and storing data in-chain for faster and more secure communications is logical.

All in all, observations showed that most papers focus more on one aspect or sub-technology of the blockchain while presenting models in their own words and not mentioning all details.

## The aim of standardization

The purpose of this paper is to provide guidelines to simplify the understanding of blockchain models generated for healthcare. Hence, we are suggesting a model with seven layers for the researchers to present their healthcare blockchain prototypes. As research shows (38), a clear description of ICOs leads to more success among users and investors. Hence, we present this paper to recommend a model to unify the prototypes. Since technology development is dependent on standards, protocols, and rules, we hope that the current study could be seen as an initiation to lead models to more efficient ones.

Software standards aim to utilize previous experience in order to form rules by developing the project to meet the future needs more comprehensively, shaping the structure to be quickly understandable, and designed in a modular way to be easily updatable. (Figure 1) Our aim was not to define a standard but to suggest study layers in representing blockchain models generated by the researchers. We firmly believe blockchain needs to have a standard for its maturation, and this can only transpire by a revolution.

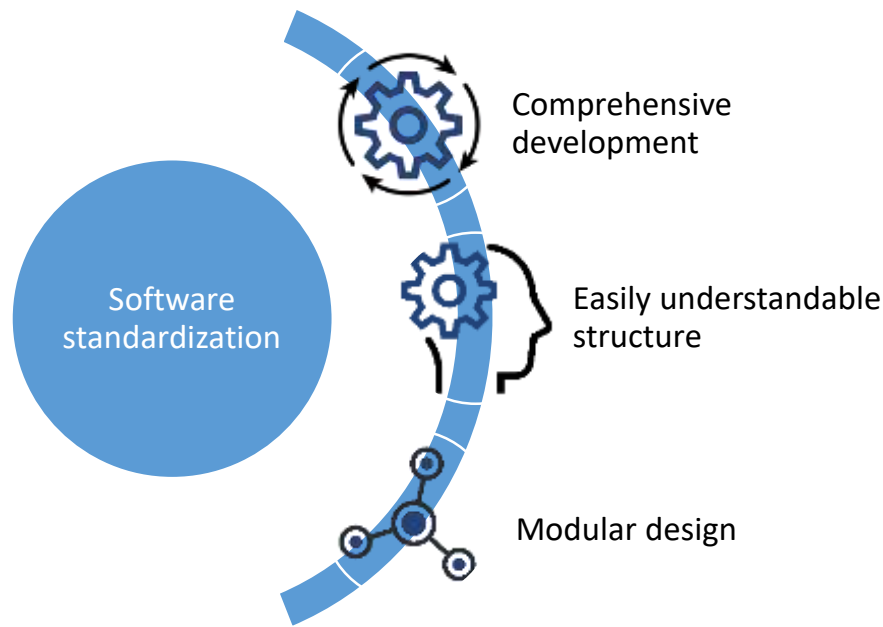

*Figure 1: The three advantages of standardization in software development.*

The current Internet protocol, known as TCP/IP, also had a similar history with blockchain technology. Bochmann et al. (39) reported the 1970s and 80s network protocols' history, OSI 7 layers and TCP/IP, by saying that we have had a long journey, many people worked, and various protocols were defined until the technology reached its current place.

One of the previous attempts toward standardization of blockchain was the Ethereum Improvement Proposal (EIP). This proposal delivers details on how to build united technical features within the community for Ethereum implementations while breaking down its standard into four levels. The *core*, for the consensus fork; *Networking* for the protocol improvements; *Interface* for the interoperability changes like API specifications, and the fourth level ERC (Ethereum Request for Comments) which offers application advances like wallet formats (40). Another practice toward it, is the "Chain Open Standard," while its developers assert that it is the pioneer in transaction speed, privacy solution, smart contract enforcement, and operational load volume (41).

In line with what was mentioned, Singhal et al. (42) suggested using a five-layer model to have a clear understanding of blockchain technology. The reason for choosing five layers was to accommodate a flexible, robust, and uncomplicated model for layering the blockchain. Those five layers are as follows, respectively: Application, Execution, Semantic, Propagation, and Consensus. Since blockchain is a meta-technology (43), the better approach for learning it is to divide the educational process into distinct layers.

## How we end up with seven layers?

Throughout a healthcare blockchain Delphi study, we asked 20 experts to identify essential blockchain qualities. This Delphi study was part of more extensive research to present a blockchain model, which we also used to inquire about our question about this specific case. We asked them which of the noted (table 1) twelve blockchain qualities are essential to consider as a segregated layer if we plan to propose the implementation of blockchain technology into divided sub-layers. We also appended that, blockchain is a meta-technology shaped by multiple computing techniques. Hence, to study blockchain models, we seek to determine which of these twelve features can be considered as an individual layer.

The experts who joined this survey were specialized in blockchain technology for at least one year and from five different countries. They selected their desired number of qualities out of these twelve features, and next, we summed up all of their votes. The experts were free to choose as many features as they liked. Seven of these features captured more than 50% of the votes. Not only did they rate these layers as the most important ones on the list, they also ranked the layers' importance. The percentage of votes might represent the significance of each layer. Hence, we are going to discuss them in order of their vote numbers. (Table 1 represents the detailed results of this survey based on their ranks)

*Table 1: List of blockchain features rated by twenty experts.*

|    | Layer Name | Number of Votes | Percentage of Experts Endorsed this | Rank |
|----|------------|-----------------|-------------------------------------|------|
| 1  | Privacy | 18 | 90% | 1 |
| 2  | Storage | 16 | 80% | 2 |
| 3  | Smart contracts | 16 | 80% | 2 |
| 4  | Distribution (Network) | 11 | 55% | 5 |
| 5  | Cryptography | 13 | 65% | 4 |
| 6  | Application | 10 | 50% | 6 |
| 7  | Consensus | 10 | 50% | 6 |
| 8  | Computation | 6 | 30% | 8 |
| 9  | Semantic (Ledger) | 6 | 30% | 8 |
| 10 | Cryptocurrency | 4 | 20% | 10 |
| 11 | Mining | 1 | 5% | 11 |
| 12 | Joining Mechanism | 1 | 5% | 11 |

# The seven layers of blockchain in healthcare

We tried to end up with fewer layers while preserving key blockchain features. Therefore, we selected the top seven layers out of the survey results. The represented blockchain traits are in terms of integrity, unity, robustness, and their impact on making blockchain an essential path to solve interoperability issues. We also focused on simplicity, dependency, and the factors that were elaborated on by previous papers. Next, regarding the diversity of blockchain designs done for healthcare, we tried to add relevant notes to clarify the general marks in each layer.

The sequence of layers represents their rank based on the number of votes and does not necessarily express their priority. We will discuss them separately, although most of the issues are satisfied for those who use the already available platforms like Hyperledger and Ethereum. In the same way, all layers are necessary to be reconsidered based on the design purpose and practice.

## *Layer 1:* Privacy

On bringing blockchain to the healthcare sector, the most crucial achievement would be protecting patient privacy. This layer is to cover the user's privacy along with healthcare ethics considering cryptocurrency techniques. Since the main aim of employing blockchain for healthcare data sharing is its underlying architecture plus solid data structure for open and secure transactions, (44) the essential prerequisite of sharing healthcare data is its security (45). Cryptography plays a vital role in ensuring the safety of access to medical data, as healthcare data has always been the primary focus of hackers (46–49). The privacy layer is responsible for safeguarding the medical data, making it available to the genuine needs for treatment and research after preserving ethics.

### Privacy and Ethics

When it comes to sharing patients' data, the main concern becomes **Ethics**. In regards to healthcare data, there are four philosophical concepts (50) that should be highlighted: autonomy, beneficence, non-malfeasance, and justice.

**Autonomy**: is the right of information self-controlling. Hence, all patients should be able to decide on their own, where/when to publish their data, while that is the primary goal of most healthcare Blockchain projects in the world.

**Beneficent**: generally means doing good and the tendency of well-being (51). The integrity of the blockchain system can meet the term. Integrity means the system works fine while blockchain comes in, primarily to satisfy that need.

**Non-Malfeasance:** do not allow harm to happen, lower risk, and increase trust (52–54). Anytime trust is reached in the healthcare system, and patients feel confident with their physicians, they would readily disclose their medical histories. Otherwise, in the course of refusing to reveal information, the case becomes more laborious; hence, hindering the treatment process (55). Consequently, trust is the foundation of non-malfeasance, which is also the main objective of Blockchain technology.

**Justice:** providing what anyone owes or deserves (52). Directly reached by healthcare interoperability, when data availability, accuracy, and security is satisfied for everyone (56). By the correct implementation of the blockchain platform, all of these can be met.

## *Layer 2:* Storage

One of the first decisions in designing a blockchain solution for healthcare would be whether to store medical data inside or outside the blockchain. This decision has a high impact on blockchain scalability, replication volume, and network speed (57). The blockchain design depends on its application, whether that is employed to share data between the same HIS (Health information system) software, connecting various software frameworks, or other purposes. Based on the use, the decision would affect the structure of the shared ledger, cryptography techniques, and even all the posterior layers of the desired blockchain.

In the case of *on-chain*, the data are encrypted and stored inside the blockchain ledger. Hence, information becomes accessible, transparent, and tamper-proof. This solution would be sufficient for use-cases with transactions small in size, identical in structure, and consistent in file type. An excellent example of this type of blockchain is cryptocurrencies such as Bitcoin and Ethereum since we are unable to store exponential data with various formats in this model.

On the other hand, healthcare data consist of various file types. MRI and X-Ray images (58,59), genomic data (60–62), structured data in software repositories (63,64), unstructured data (65), such as various kinds of reports and forms, numerous paged scanned patients' documents, etc. All of these types mentioned above shaped the medical data, which leads healthcare to a homomorphic data structure. Maybe the most significant difference between healthcare and cryptocurrency blockchain.

In the case of *off-chain*, the ledger would be the place to store the permission tables and transaction data logs. Ethically, logs must clarify who has accessed the records of a patient at a specific time or place. As data are getting out of the secure environment of the blockchain, the solution should surpass by solving the single point of failure for medical histories and files. Moreover, developing a new mechanism is required for querying the data. In the off-chain model, there should be a security plan to store file addresses/pointers, data hash (66), and even some meta-data, on the ledger or smart contracts, while accessible by all authorized users.

Both on-chain and off-chain have their *pros and cons* and should be selected based on the case purpose. In terms of on-chain, we have more immutability, availability, transparency with the cost of high computation needs, data redundancy, and information revealing. Hence, the network becomes slow, complex, expensive, and hard to work with. When it comes to off-chain, the system can be more scalable, transferring files of bigger-size and various types becomes more comfortable, and medical histories can be kept private. However, the design solution should not compromise blockchain benefits. (57) Table 2 outlines both solutions.

Note that ledger is also discussed here as a part of the storage system, although experts deprioritize it as a single layer. In both on-chain and off-chain cases, the ledger is transparent, tamper-proof, and can be shared among all nodes.

*Table 2: The pros and cons of the on-chain and off-chain model. The gray shading shows the advantage, and the other ones are considered as disadvantages. Note that solutions should be developed in a way to overcome obstacles.*

| Layer 1 | **On-chain** | **Off-chain** |
|---:|---|---|
| *Cost* | Higher cost of data redundancy and computations. | Lower disk space and computations lead to lower costs. |
| *Time and speed* | It consumes more time to compute every transaction and synchronize all nodes. | One or a few providers keep a copy of the data. Hence, less time is consumed for synchronization. |
| *Computations* | Every transaction needs computations. | The data are queried for patient treatment; hence, computations are needed only at the time of data generation and fetching. |
| *Replication* | All full-nodes replicate the data. | Based on the solution, one or a few providers keep the data. |
| *Privacy* | Data is revealed for computations. | Data is kept by the providers until a physician request to access patient history. |
| *Immutability* | Information is in the blocks. Hence, no one can alter it without having at least 51 percent of the network power. | One or a few mediums keep the data. |
| *Scalability* | Nodes need more resources to join the system. | Easier scalability features. |
| *Data availability* | All full-nodes are available to reply to the information requested. | The data provider should be accessible. |

The primary database of blockchain is the shared **ledger** that brings trust and transparency. The ledger is a key characteristic of blockchain that establishes clarity between untrusted parties. It can be a solution for various healthcare stakeholders to join a network in the absence of trust and transfer valuable data while records are safeguarded, and transactions are apparent to all.

Ledger is the database presented to all. In centralized networks, the databases can only be accessed by the company's administrators, but in a decentralized one, it is available to all users. Blockchain is a new trend in trust and security. In centralized networks, security was to hide servers and protect them by firewalls, but in the decentralized blockchain, transparency and distribution provides the safety.

The ledger is an append-only database of transactions. Blockchain distributes mirror copies of the ledger through the peer-to-peer network after new transactions are done and accepted. Hence, structuring the ledger contents should make the whole system a more trustable and transparent place. In figure 2, we depict a hypothetical transaction in a healthcare blockchain system.

Page **8** of **24**

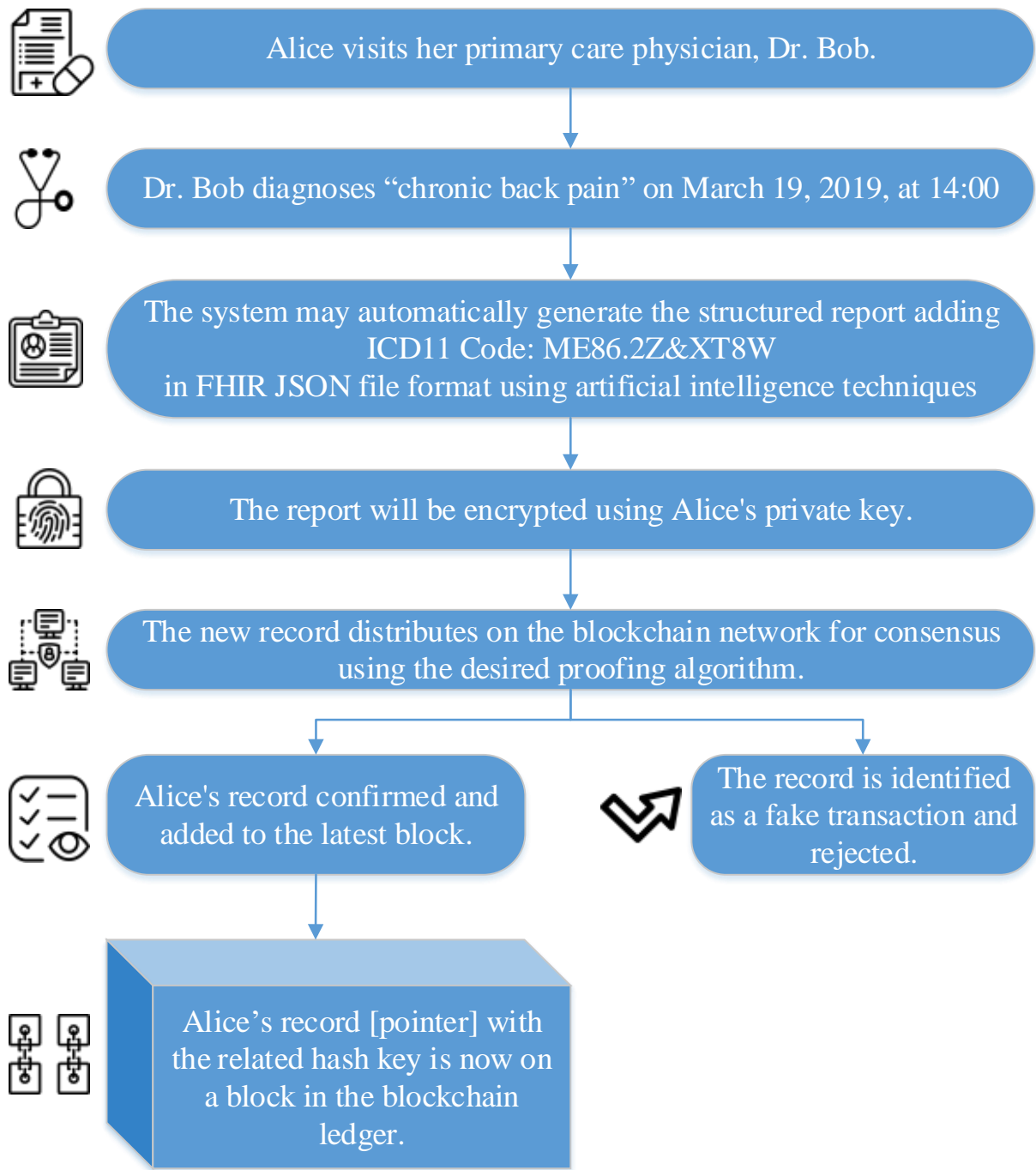

*Figure 2: A hypothetical transaction in a healthcare blockchain*

## *Layer 3:* Smart Contract (Execution)

The unique feature of blockchain technology "smart contracts" is defined here by considering that the current layer adds real computational power to the blockchain, making it a prominent system. Smart contracts are codes that enable blockchain to transfer more than just values or coins; in addition, setting policies, rules, and regulations. The emergence of asset transfer in blockchain has made engineers utilize it in other settings such as healthcare.

Smart contracts are determined as self-executable codes that run after certain conditions are met. Those codes are written based on agreements or contracts between mistrusting parties. A clear example could be the patient's data exchange between healthcare providers and insurance companies to pay back their share of treatment costs. Accordingly, that would facilitate the complicated bureaucracy processes by reducing the costs of administration and converting paper-based contracts and procedures into easy byte-codes.

By the time of submission and after smart contract codes are stored on the blockchain, they cannot be revoked. Hence, it is imperative to test and be sure of the codes' functionality by overcoming any ambiguity before deploying the codes. It is essential to develop smart contracts in an auditable manner for verification purposes and to log all events (67) during the running time. Most healthcare policies - in the format of smart contracts - are regulated for long-term resolutions.

Blockchain is not a data repository; besides, smart contracts once deployed are permanent and immutable. Therefore, to allocate resources in a better way, variables should be developed after categorizing them into two permanent and transient values to store vital information permanently in storage, and the temporary ones in memory.

Smart contracts are a digitized formation of confidential agreements among untrusted parties, which will be the next-generation business paradigm.

In parallel with healthcare ethics, patient privacy, and access control mechanism, the execution should be running while considering permission tables to satisfy data access ethics.

## *Layer 4:* Distribution (Network)

The decentralization has a meaning here, where no central authority controls the whole network. Blockchain is a network of nodes joined together while they are not limited to synchronize to a centralized server. A node finds the nearest peer and gets the latest transaction. It also checks hashes to ensure that the received ledger is the one that the majority of network members agree to be true.

Decentralization is what safeguards networks against vulnerabilities like a single point of failure. By discarding the central server and distributing resources, there is no single point for hackers to target. An attack on a user will only suspend that specific node while other peers continue using the network by pairing to different available routes.

This is how blockchain makes a cohesive network, and a hostile person needs to own 51% of the nodes to spoil the system. Distributing resources can overcome vulnerabilities like Denial of Service (DOS) types of attacks that are prevalent on the web. The bigger the network, the more powerful it is, and the harder it would be to down the system.

The developer of decentralized apps has to address solutions for some concerns. Such as:

- How peers are going to find each other
- Where are the nearest nodes
- How blocks are propagated and forwarded (to ensure all the network is synchronized with the latest and most trustable ledger)
- How new nodes join the network
- How to overcome the nodes latency issue
- How the route tables can lead a node to the nearest available host, especially for off-chain cases

## *Layer 5:* Cryptography

Cryptography is what brings an immutable structure to the blockchain. The encryption algorithms formed the basis of blockchain by their principles. Regarding the details of these principles, it could be a trade-off as sometimes improving the security of the blockchain is a loss of another feature. This trade-off implies that we have to compromise on features that are desirable yet are opposite. (68) Next, data in healthcare has various scopes; hence, decrypting it is time-consuming, which demands solutions.

Applying the **digital signature** as a strategic blockchain technology feature for healthcare has some cons that have to be considered (69,70). As an example, barriers can be key management or reducing the speed of access to emergency medical data. Key management and confiding the private key is an issue that has to be considered since some patients might lack access to smartphones (67,71). Tian et al. (72) simulated a model for accessing the patient keys in case of emergency to make medical data available in the blockchainized world.

The flip side to this coin is the emergence of **quantum computers** (73,74) and their high potential for computing. This processing power threatens the blockchain digital signature cryptography integrity (75). Although we are far from quantum days (75), systems are much better designed in a way that would be able to change their cryptography algorithms by employing post-quantum ones when needed (76,77).

## *Layer 6:* Application

The decentralized software engineered for blockchain should be capable of working on a peer-to-peer network, in comparison to regular applications that mostly connect to a particular data server. Hence, the appropriate development of a framework based on the setting's requirement is selected here. The scalability of the client-side app, the modular-based programming, and API (Application Programming Interface) connection, will reach a decision here.

The client-side medical software has to be developed more cautiously since any corruption in the codes might lead to a patient's harm (78). A study (79) shows that at least 173 million records of health data were breached since late 2009 in the US. While it is vital to have instant access to health information, the weaknesses can lead to severe social and ethical problems caused by malware, data breaches, or hackers attacking the system (80,81).

## *Layer 7:* Consensus

What is most remarkable about any blockchain system is its consensus type. The consensus is the way that any untrusted peer agrees to add a new transaction in the decentralized network. It can be respected as the most crucial game-changer in any blockchain as system scalability and performance is dependent on it. The consensus is responsible for defining how a peer can be trusted, in addition to how we chain a new transaction to the previous ones.

Its primary duties are to preserve ownership and maintain system consistency. To address these qualities, it has to incentivize some users to collaborate in system integrity. Blockchain strategy invites some nodes to participate in solving some mathematical computations to distinguish honest from dishonest nodes. Nevertheless, each system requires its specific consensus, depending on the need for scalability, availability, and consistency. There are many consensus algorithms available, while some might also develop their own.

Liu et al. (82) investigated blockchain game theories and introduced the mechanism as follows:

- A node announces a new transaction.
- Other nodes verify the honesty of both the node and its current transaction.
- A bundle of transactions is broadcasted to be verified by the network.
- Some nodes participate in the process of validating this new block by running blockchain consensus functions.
- The new block is chained to the previous ones, and nodes synchronize their local ledger by replicating the novel updates.

Sonnino et al. (83) believe that blockchain has scalability issues for large-scale usage, while a proper consensus can fix this issue. That is why many seek alternative solutions acknowledging that building a new consensus demands a deep understanding of other protocols. Following, they classified protocols by ten features and analyzed them using the same framework. They recalled three points of classical consensus (84) that for all faultless processors, a value is assigned. Next, all come to one identical value, and lastly, if all processors submit the same input, then that particular value is validated. Subsequently, they distinguish the blockchain consensus and compared several protocols to mark their key features and settings that they work in. Another consensus study is what Wang et al. (85) did on the permissionless consensus mechanisms. They underlined how blockchain could impact different fields by explaining the purpose of blockchain in these areas while mentioning how blockchain consensus works comprehensively.

Another group of scientists (86) believes that the consensus type is the crucial point in blockchain performance and scalability. They claimed that they could simplify the blockchain consensus protocols to evaluate their performance, security, and design properties to answer any future concerns.

The other concern is what if 51% of the nodes are in collusion to impair the system integrity. Schrepel (87) attempted to answer this question in his comprehensive article. He analyzed collusive agreements in various scenarios considering suspicious acts by the users, developers, or miners, either in consensus mechanism, smart contracts, public, or private blockchain. Then, he addressed the situations and highlighted how we could efficiently exit this challenge by using smart contracts. Following, he stated how laws and regulations could also help to prohibit trust between colluders and reach a safe, distributed environment.

After all, we suggest that precise attention required for the process of developing a consensus protocol for a system. The design and implementation of blockchain consensus profoundly impact the expansion of the system and its wide-scale extensibility. Moreover, consensus influences the speed, security, and system operation while also attracting nodes to associate in safeguarding the network by its incentivization craft. Hence, any consensus type by its precise features belongs to the related areas and has its unique outcomes.

## Functional Synthesis

We have proposed to study healthcare blockchain models in seven layers to consider features of the technology as significant traits. Several technologies join in shaping the blockchain and making it an outstanding solution to overcome medical data-sharing issues. Each technique, regardless of their priority, has various options to be considered. Therefore, we suggest studying blockchain models in separate levels to emphasize all of their model features to achieve the best applicable proposal.

Despite the fact that blockchain has made swift improvements in finance and supply chain, it is still immature in healthcare (88), whereas it has to address different solutions for the medical setting. As an example, the ledger in cryptocurrencies has an invariant structure, while blockchain in healthcare has to cope with various forms. Moreover, it has to resolve multi-type data with diverse structures, usually located in distinct hosts.

Blockchain, with its modular characteristic, commits its integrity, immutability, and flexibility to be employed in different applications in healthcare such as patient-driven interoperability (11,89), privacy-preserving data exchange (90), and improving clinical research (91). Hence, if developed appropriately, it can lead to a revolution in healthcare.

The "Data sharing" term is defined by the possibility of providing data for various applications or users. In healthcare, more explanation is needed to satisfy continuity of care. The time that we can claim we have "data sharing" in our health system is when the data would be available to the authenticated person as an aggregate, accurate, and complete entity. Blockchain aims its capability to satisfy this need, yet on a premise that all healthcare stakeholders discover this potential by actively employing it in their systems.

Mackey et al. (88) believe that the modularity of blockchain enables it to adapt to multiple usages in the health sector. Following, they discussed six questions to verify the possibility of this approach; hence, they used this procedure to set some examinations and outlined how a future healthcare blockchain would work, based on the currently available resources. They gathered cross-disciplinary specialists to discuss various applications of blockchain in healthcare while they emphasized the essential requirement to take this cutting-edge technology forward as an industry standard. Finally, they asserted that newly raised uncertainties on the blockchain, such as interoperability of legacy systems, consensus-built credibility, and rising open platforms, can be addressed by the technical standard.

What makes blockchain a robust system, is its well-combined techniques to change an untrusted network to a fully trusted one while it is necessary for presented models to clarify how they apply those techniques in their model. We highlighted these features to ask system architects to make their ideas toward these points more transparent. There are seven principal characteristics to be considered when a blockchain model is going to be presented; Two for healthcare; and five for technical aspects. As mentioned, the first two are for interpreting the system behavior toward medical files, and how privacy is preserved. The further five approaches are as follows:

First, smart contracts, second, determining how nodes route in the network and detail of how the ledger would be distributed. Third, the policies toward encryption. Fourth, the front-end interface where users work with the system, and fifth, the consensus type and the method of incentivizing miners for their commitment to addressing the system integrity. That is the way we believe readers might better obtain insight into the whole model presented.

The medical data stands out when it ends up in a unified reality rather than a fragmented one. Blockchain can reshape healthcare by transforming traditional communication problems with the enforcement of blockchain networks (85). However, blockchain has many hurdles to being adopted widely in legacy medical software and needs to be matured in order to become reliable. Also, more research is warranted to regulate and address best practices for healthcare (92).

## Conclusion

Blockchain is a joined of multiple techniques that lead to a safe, transparent, and privacy-preserving infrastructure. The design of this meta-technology demands for architectural decision subjects like confidential data, storage methods, regulations, block propagation, encryption algorithms, interfaces, and consensus mechanisms. That is why we request the separation of these concerns in the design process for more clarity.

In this study, we recommended system architects to explain their models giving detailed information in these seven layers. We suggested abstracts that would shed more light on the model presentations for healthcare blockchain. This particular model will improve academic voice in presenting healthcare blockchain models and highlight each essential blockchain feature individually. These presented features are seven traits of blockchain, discussed as segregated layers. The application and details for each one can be studied more to enhance blockchain utilization in healthcare; that is how we conceived there is a lack of standardization for it.

Bringing standards in technologies can help it to become scalable by creating many opportunities. The standard or somehow regulations in healthcare blockchain can lead to wide-scale adoption, resulting in enhancement of clinical research, empowerment of patient control of self-data, and the emergence of precision medicine. All these can give rise to health services and advance social wellness.

Overall, we proposed this layered approach to the implementation of blockchain in the healthcare sector and explored fundamental concerns in developing this technology to point out each as a layer. We believe describing models in this way can improve the development of blockchain in healthcare systems. Therefore, let the revolution begin!

## Limitations

The major limitation of the study is that blockchain, a cutting edge technology, is relatively new, although ten years have passed since its introduction. However, few projects have been carried out in healthcare, and a small number of outstanding research papers are available in this setting; moreover, there is a lack of expert knowledge, so we get together needed information mostly based on each layer.

## Future studies

The topic of the paper is still new and surely can be enhanced more. Hence blockchain experts are appreciated to share their experience and address their findings to advance this issue. Even a council can be shaped to standardize the conversation in blockchain technology, or a group of professionals may debate the regulatory requirements.

## Acknowledgment

The present article was adopted from Amir Hossein Zolfaghari's M.Sc. thesis in Medical informatics. The authors would like to thank the Research Vice-Chancellor of Shiraz University of Medical Sciences for financially supporting the research (Contract No. 97-01-07-16810 and the research ethics certificate ID: IR.SUMS.REC.1397.863, that can be verified on this [link)](). The authors wish to thank Mr. H. Argasi at the Research Consultation Center (RCC) of Shiraz University of Medical Sciences for his invaluable assistance in editing this manuscript.

## Conflict of Interest

The authors state that they have no conflicts of interest to declare.

# References

1. Nakamoto S. Bitcoin: A Peer-to-Peer Electronic Cash System. [cited 2018 Mar 30]; Available from: www.bitcoin.org

2. Swan M, de Filippi P. Toward a Philosophy of Blockchain: A Symposium: Introduction. Metaphilosophy [Internet]. 2017 Oct [cited 2019 Feb 7];48(5):603–19. Available from: http://doi.wiley.com/10.1111/meta.12270

3. Pilkington M. Blockchain Technology: Principles and Applications. Res Handb Digit Transform [Internet]. 2015 Sep 18 [cited 2019 Jan 27];1–39. Available from: https://papers.ssrn.com/sol3/papers.cfm?abstract_id=2662660

4. Narayanan, A., J. Bonneau, E. Felten, A. Miller and SG. Bitcoin and Cryptocurrency Technologies: A Comprehensive Introduction. [Foreword: The Long Road To Bitcoin]. Princet Univ Press [Internet]. 2016 [cited 2019 Jan 28];304. Available from: https://press.princeton.edu/titles/10908.html

5. Fisch C. Initial Coin Offerings (ICOs) to Finance New Ventures: An Exploratory Study. SSRN Electron J [Internet]. 2018 Sep 29 [cited 2019 Jan 28]; Available from: https://www.ssrn.com/abstract=3147521

6. Radanović I, Likić R. Opportunities for Use of Blockchain Technology in Medicine. Appl Health Econ Health Policy [Internet]. 2018 Oct 18 [cited 2019 Jan 28];16(5):583–90. Available from: http://www.ncbi.nlm.nih.gov/pubmed/30022440

7. Tung JK, Nambudiri VE. Beyond Bitcoin: potential applications of blockchain technology in dermatology. Br J Dermatol [Internet]. 2018 Oct [cited 2019 Jan 28];179(4):1013–4. Available from: http://www.ncbi.nlm.nih.gov/pubmed/29947078

8. Roman-Belmonte JM, De la Corte-Rodriguez H, Rodriguez-Merchan EC. How blockchain technology can change medicine. Postgrad Med [Internet]. 2018 May 19 [cited 2019 Jan 28];130(4):420–7. Available from: http://www.ncbi.nlm.nih.gov/pubmed/29727247

9. Angraal S, Krumholz HM, Schulz WL. Blockchain Technology; Applications in Health Care. Circ Cardiovasc Qual Outcomes [Internet]. 2017 Sep [cited 2019 Jan 28];10(9). Available from: http://www.ncbi.nlm.nih.gov/pubmed/28912202

10. Funk E, Riddell J, Ankel F, Cabrera D. Blockchain Technology: A Data Framework to Improve Validity, Trust, and Accountability of Information Exchange in Health Professions Education. Acad Med [Internet]. 2018 Dec [cited 2019 Jan 28];93(12):1791–4. Available from: http://www.ncbi.nlm.nih.gov/pubmed/29901658

11. Gordon WJ, Catalini C. Blockchain Technology for Healthcare: Facilitating the Transition to Patient-Driven Interoperability. Comput Struct Biotechnol J [Internet]. 2018 [cited 2018 Nov 18];16:224–30. Available from: http://www.ncbi.nlm.nih.gov/pubmed/30069284

12. Summerton N. The medical history as a diagnostic technology [Internet]. Vol. 58, British Journal of General Practice. 2008 [cited 2018 Feb 24]. p. 273–6. Available from: http://www.ncbi.nlm.nih.gov/pubmed/18387230

13. Eliav E. The importance of collecting patients' medical histories. Quintessence Int [Internet]. 2012 Feb;43(2):91. Available from: http://www.ncbi.nlm.nih.gov/pubmed/22257868

14. Mortazavi H, Rahmani A, Rahmani S. Importance, Advantages, and Objectives of Taking and Recording Patient's Medical History in Dentistry. Int J Med Rev Vol 2, No 3 Summer 2015 [Internet]. 2015;2(3):287–90. Available from: http://journals.bmsu.ac.ir/ijmr/index.php/ijmr/article/view/112

15. Muhrer JC. The importance of the history and physical in diagnosis. Nurse Pract [Internet]. 2014 Apr 13 [cited 2018 Feb 24];39(4):30–5. Available from: http://www.ncbi.nlm.nih.gov/pubmed/24584168

16. Ghosh D, Karunaratne P. The importance of good history taking: a case report. J Med Case Rep [Internet]. 2015 Apr 30 [cited 2018 Feb 24];9:97. Available from: http://www.ncbi.nlm.nih.gov/pubmed/25924859

17. Menachemi N, Collum TH. Benefits and drawbacks of electronic health record systems. Risk Manag Healthc Policy [Internet]. 2011 May [cited 2018 Jan 30];4:47–55. Available from: http://www.ncbi.nlm.nih.gov/pubmed/22312227

18. Jones SS, Adams JL, Schneider EC, Ringel JS, Mcglynn EA. Electronic Health record Adoption and Quality Improvement in US Hospitals. Am J Manag Care [Internet]. 2010 Dec [cited 2018 Feb 18];16(12 Suppl HIT):64–71. Available from: http://www.ncbi.nlm.nih.gov/pubmed/21314225

19. Walker J, Meltsner M, Delbanco T. US experience with doctors and patients sharing clinical notes. BMJ [Internet]. 2015 Feb 10;350(February):g7785. Available from: http://www.bmj.com/lookup/doi/10.1136/bmj.g7785

20. NHS England. Better information means better care. NHS [Internet]. 2014;2. Available from: https://www.england.nhs.uk/wp-content/uploads/2014/01/cd-leaflet-01-14.pdf

21. Buntin MB, Burke MF, Hoaglin MC, Blumenthal D. The Benefits Of Health Information Technology: A Review Of The Recent Literature Shows Predominantly Positive Results. Health Aff [Internet]. 2011 Mar 1;30(3):464–71. Available from: http://content.healthaffairs.org/cgi/doi/10.1377/hlthaff.2011.0178

22. Wang H, Chen K, Xu D. A maturity model for blockchain adoption. Financ Innov [Internet]. 2016;2(1):12. Available from: http://jfin-swufe.springeropen.com/articles/10.1186/s40854-016-0031-z

23. Peck ME. Blockchain world - Do you need a blockchain? This chart will tell you if the technology can solve your problem. IEEE Spectr. 2017;54(10):38–60.

24. Wust K, Gervais A. Do you Need a Blockchain? 2018 Crypto Val Conf Blockchain Technol [Internet]. 2018;(i):45–54. Available from: https://ieeexplore.ieee.org/document/8525392/

25. Manish Shanbhag. Do you need a BlockChain Solution? [Internet]. Linkedin - https://www.linkedin.com/pulse/do-you-need-blockchain-solution-manish-shanbhag/. 2019 [cited 2019 Jan 30]. Available from: https://www.linkedin.com/pulse/do-you-need-blockchain-solution-manish-shanbhag/

26. Shekelle PG, Morton SC, Keeler EB. Costs and benefits of health information technology. Evid Rep Technol Assess (Full Rep) [Internet]. 2006 Apr [cited 2019 Jan 31];(132):1–71. Available from: http://www.ncbi.nlm.nih.gov/pubmed/17627328

27. Reisman M. EHRs: The Challenge of Making Electronic Data Usable and Interoperable. P T [Internet]. 2017 Sep [cited 2019 Jan 31];42(9):572–5. Available from: http://www.ncbi.nlm.nih.gov/pubmed/28890644

28. Aspden P, Corrigan JM, Wolcott J, Erickson SM. Patient Safety:: Achieving a New Standard for Care Editors, Committee on Data Standards for Patient Safety [Internet]. Vol. 550. 2004. 0–309 p. Available from: http://www.nap.edu/catalog/10863.html

29. Edwards RA, Venugopal S, Navedo D, Ramani S. Addressing needs of diverse stakeholders: Twelve tips for leaders of health professions education programs. Med Teach [Internet]. 2017;0(0):1–7. Available from: https://doi.org/10.1080/0142159X.2017.1396307

30. Krawiec R, Barr D, Killmeyer J, Filipova M, Quarre F, Nesbitt A, et al. Blockchain : Opportunities for Health Care. NIST Work Blockchain Healthc [Internet]. 2016;(August):1–12. Available from: https://www2.deloitte.com/content/dam/Deloitte/us/Documents/public-sector/us-blockchain-opportunities-for-health-care.pdf

31. Nichol PB, Brandt J. Co-Creation of Trust for Healthcare: The Cryptocitizen Framework for Interoperability with Blockchain. ResearchGate [Internet]. 2016;(August):0–9. Available from: https://www.researchgate.net/profile/Peter_Nichol2/publication/306013124_Co-Creation_of_Trust_for_Healthcare_The_Cryptocitizen_Framework_for_Interoperability_with_Blockchain/links/57aa227808ae7a6420bcd091/Co-Creation-of-Trust-for-Healthcare-The-Cryptociti

32. Zhang P, White J, Schmidt DC, Lenz G, Rosenbloom ST. FHIRChain: Applying Blockchain to Securely and Scalably Share Clinical Data. Comput Struct Biotechnol J [Internet]. 2018;16:267–78. Available from: https://doi.org/10.1016/j.csbj.2018.07.004

33. Karen B. DeSalvo, Galvez E. Connecting Health and Care for the Nation: A Shared Nationwide Interoperability Roadmap - Final Version 1.0. ONC Rep [Internet]. 2015;1–9. Available from: https://www.healthit.gov/sites/default/files/nationwide-interoperability-roadmap-draft-version-1.0.pdf%0Ahttp://www.healthit.gov/sites/default/files/nationwide-interoperability-roadmap-draft-version-1.0.pdf

34. hl7.org. FHIR Release 3 (STU) [Internet]. www.hl7.org/fhir/. 2017 [cited 2018 Sep 1]. Available from: https://www.hl7.org/fhir/

35. Bender D, Sartipi K. HL7 FHIR: An agile and RESTful approach to healthcare information exchange. In: Proceedings of CBMS 2013 - 26th IEEE International Symposium on Computer-Based Medical Systems [Internet]. IEEE; 2013 [cited 2018 Sep 1]. p. 326–31. Available from: http://ieeexplore.ieee.org/document/6627810/

36. Zhang A, Lin X. Towards Secure and Privacy-Preserving Data Sharing in e-Health Systems via Consortium Blockchain. J Med Syst. 2018;42(8).

37. Zhou L, Wang L, Sun Y. MIStore: a Blockchain-Based Medical Insurance Storage System. J Med Syst [Internet]. 2018 Jul 2 [cited 2018 Dec 3];42(8):149. Available from: http://www.ncbi.nlm.nih.gov/pubmed/29968202

38. Stanley M. The Application of Behavioural Heuristicsto Initial Coin Offerings Valuation and Investment. J Br Blockchain Assoc [Internet]. 2019 May 4;2(1):1–7. Available from: https://jbba.scholasticahq.com/article/7776-the-application-of-behavioural-heuristics-to-initial-coin-offerings-valuation-and-investment

39. Bochmann G V., Rayner D, West CH. Some notes on the history of protocol engineering. Comput Networks [Internet]. 2010;54(18):3197–209. Available from:

http://dx.doi.org/10.1016/j.comnet.2010.05.019

40. Johnson N, Entriken W, Jameson H, Beregszaszi A, Savers N, Hirai Y, et al. What is an EIP? [Internet]. GitHub. Available from: https://github.com/ethereum/EIPs/blob/master/EIPS/eip-1.md

41. Chain.com. Chain Open Standard for Blockchain [Internet]. Chain Inc. Available from: https://chain.com/press-releases/chain-and-global-financial-firms-unveil-open-standard-for-blockchain/

42. Singhal B, Dhameja G, Panda PS. Beginning Blockchain; A Beginner's Guide to Building Blockchain Solutions [Internet]. Berkeley, CA: Apress; 2018 [cited 2018 Dec 30]. 386 p. Available from: http://link.springer.com/10.1007/978-1-4842-3444-0

43. Mougayar W. Working as a layer on top of the internet, blockchain is an instrument of change. LSE Bus Rev [Internet]. 2016;1–3. Available from: https://blogs.lse.ac.uk/businessreview/2016/07/07/working-as-a-layer-on-top-of-the-internet-blockchain-is-an-instrument-of-change/

44. Randall D, Goel P, Abujamra R. Blockchain Applications and Use Cases in Health Information Technology. J Heal Med Informatics [Internet]. 2017;08(03):8–11. Available from: https://www.omicsonline.org/open-access/blockchain-applications-and-use-cases-in-health-information-technology-2157-7420-1000276.php?aid=91911

45. Bosu A, Iqbal A, Shahriyar R, Chakroborty P. Understanding the Motivations, Challenges and Needs of Blockchain Software Developers: A Survey [Internet]. 2018. Available from: http://arxiv.org/abs/1811.04169

46. Topics KEY. McAfee Labs Threat Report The WannaCry malware attack infected. 2017;(September):1–66. Available from: https://www.mcafee.com/ca/resources/reports/rp-quarterly-threats-sept-2017.pdf

47. Luana PASCU. Healthcare is top target for hackers; 1 in 3 records to be breached in 2018, study says [Internet]. bitdefender. 2017 [cited 2018 Mar 30]. Available from: https://www.bitdefender.com/box/blog/iot-news/healthcare-top-target-hackers-1-3-records-breached-2018-study-says/

48. David Wagner ZC. Why Healthcare Is A Top Target For Hackers [Internet]. healthitoutcomes.com. 2018 [cited 2018 Mar 30]. Available from: https://www.healthitoutcomes.com/doc/why-healthcare-is-a-top-target-for-hackers-0001

49. Mearian L. Cyberattacks will compromise 1-in-3 healthcare records next year | Computerworld [Internet]. 2015 [cited 2018 Mar 30]. Available from: https://www.computerworld.com/article/3013013/healthcare-it/cyberattacks-will-compromise-1-in-3-healthcare-records-next-year.html

50. Goodman KW, Cambridge University Press., Cambridge. Ethics, Computing, and Medicine : Informatics and the Transformation of Health Care [Internet]. Goodman KW, editor. Cambridge: Cambridge University Press; 1997 [cited 2018 Oct 19]. 196 p. Available from: http://ebooks.cambridge.org/ref/id/CBO9780511585005

51. Frankena WK. Beneficence/benevolence. Soc Philos Policy Found 1987 - Univ Cambridge Press [Internet]. 1987 Mar 13 [cited 2018 Oct 20];4(2):1–20. Available from: http://www.journals.cambridge.org/abstract_S0265052500000510

52. Gracyk T. Four fundamental principles of ethics [Internet]. Minnesota State University

Moorhead. 2012 [cited 2018 Oct 20]. Available from: http://web.mnstate.edu/gracyk/courses/phil 115/Four_Basic_principles.htm

53. Www.themedicportal.com. Medical Ethics Explained: Non-Maleficence - The Medic Portal [Internet]. The Medic Portal. 2016 [cited 2018 Oct 20]. Available from: https://www.themedicportal.com/blog/medical-ethics-explained-non-maleficence/

54. Gillon R. "Primum non nocere" and the principle of non-maleficence. Br Med J (Clin Res Ed) [Internet]. 1985 Jul 13 [cited 2018 Oct 20];291(6488):130. Available from: http://www.ncbi.nlm.nih.gov/pubmed/3926081

55. Olaronke Ganiat I, Janet Olusola O. Ethical Issues in Interoperability of Electronic Healthcare Systems. Commun Appl Electron [Internet]. 2015 May 26 [cited 2018 Oct 19];1(8):12–8. Available from: http://caeaccess.org/research/volume1/number8/cae-1626.pdf

56. Almacen M. EHR Interoperability: Legal, Ethical, and Social Challenges [Internet]. Northwestern University. Northwestern University Evanston, Illinois; 2013. Available from: http://mariaa-northwesternu.weebly.com/uploads/3/1/7/1/31716081/project_3_almacenm_407dl_march2013_final.docx

57. Eberhardt J, Tai S. On or Off the Blockchain? Insights on Off-Chaining Computation and Data. In Springer International Publishing; 2017 [cited 2019 Feb 11]. p. 3–15. Available from: http://link.springer.com/10.1007/978-3-319-67262-5_1

58. Fatehi M, Safdari R, Ghazisaeidi M, Jebraeily M, Habibi-Koolaee M. Data Standards in Tele-radiology. Acta Inform Med [Internet]. 2015 Jun [cited 2018 Dec 30];23(3):165–8. Available from: http://www.ncbi.nlm.nih.gov/pubmed/26236084

59. Larobina M, Murino L. Medical image file formats. J Digit Imaging [Internet]. 2014 Apr [cited 2018 Dec 30];27(2):200–6. Available from: http://www.ncbi.nlm.nih.gov/pubmed/24338090

60. Grishin D, Obbad K, Estep P, Quinn K, Zaranek SW, Zaranek AW, et al. Accelerating Genomic Data Generation and Facilitating Genomic Data Access Using Decentralization , Privacy-Preserving Technologies and Equitable Compensation. Blockchain Healthc Today [Internet]. 2018 Dec 19 [cited 2018 Dec 30];1(0):1–23. Available from: https://blockchainhealthcaretoday.com/index.php/journal/article/view/34

61. TA B. Chapter 7, Understanding a Genome Sequence. In: Genomes [Internet]. Wiley-Liss; 2002 [cited 2018 Dec 30]. Available from: https://www.ncbi.nlm.nih.gov/books/NBK21136/

62. Ostell J. What's in a Genome at NCBI ? In: The NCBI Handbook [Internet]. National Center for Biotechnology Information (US); 2013 [cited 2018 Dec 30]. p. 3–8. Available from: https://www.ncbi.nlm.nih.gov/books/NBK169442/

63. Wade TD. Traits and types of health data repositories. Heal Inf Sci Syst [Internet]. 2014 [cited 2018 Dec 30];2(1):4. Available from: http://www.ncbi.nlm.nih.gov/pubmed/25825668

64. Ateya MB, Delaney BC, Speedie SM. The value of structured data elements from electronic health records for identifying subjects for primary care clinical trials Healthcare Information Systems. BMC Med Inform Decis Mak [Internet]. 2016 Jan 11 [cited 2018 Dec 30];16(1):1. Available from:

http://www.ncbi.nlm.nih.gov/pubmed/26754574

65. Polnaszek B, Gilmore-Bykovskyi A, Hovanes M, Roiland R, Ferguson P, Brown R, et al. Overcoming the challenges of unstructured data in multisite, electronic medical record-based abstraction. Med Care [Internet]. 2016 [cited 2018 Dec 30];54(10):e65–72. Available from: http://www.ncbi.nlm.nih.gov/pubmed/27624585

66. Eberhardt J, Heiss J. Off-chaining Models and Approaches to Off-chain Computations. In: Proceedings of the 2nd Workshop on Scalable and Resilient Infrastructures for Distributed Ledgers - SERIAL'18 [Internet]. New York, New York, USA: ACM Press; 2018 [cited 2019 Feb 11]. p. 7–12. Available from: http://dl.acm.org/citation.cfm?doid=3284764.3284766

67. Griggs KN, Ossipova O, Kohlios CP, Baccarini AN, Howson EA, Hayajneh T. Healthcare Blockchain System Using Smart Contracts for Secure Automated Remote Patient Monitoring. J Med Syst [Internet]. 2018 Jul 6 [cited 2019 Apr 11];42(7):130. Available from: http://www.ncbi.nlm.nih.gov/pubmed/29876661

68. O'Donoghue O, Vazirani AA, Brindley D, Meinert E. Design choices and trade-offs in health care blockchain implementations: Systematic review. J Med Internet Res. 2019;21(5):1–13.

69. Al. FHRF et. Digital signature and electronic certificates in health care. Stud Heal Technol Informatics - IOS Press [Internet]. 2004;110:87–9. Available from: http://ebooks.iospress.nl/volumearticle/10017

70. Zuckerman AE. Restructuring the electronic medical record to incorporate full digital signature capability. Proceedings AMIA Symp [Internet]. 2001 [cited 2019 Apr 11];791–5. Available from: http://www.ncbi.nlm.nih.gov/pubmed/11825294

71. Li P, Nelson SD, Malin BA, Chen Y, Chen Y. DMMS: A Decentralized Blockchain Ledger for the Management of Medication Histories. Blockchain Healthc Today [Internet]. 2018 Dec 31 [cited 2019 Jan 14];2(0):1–15. Available from: https://blockchainhealthcaretoday.com/index.php/journal/article/view/38

72. Tian H, He J, Ding Y. Medical Data Management on Blockchain with Privacy. J Med Syst [Internet]. 2019 Feb 3;43(2):26. Available from: http://link.springer.com/10.1007/s10916-018-1144-x

73. Michael A. Nielsen ILC. Quantum Computation and Quantum Information [Internet]. Cambridge University Press; 2011 [cited 2019 Feb 17]. Available from: https://dl.acm.org/citation.cfm?id=1972505

74. Kaye P, Laflamme R, Mosca M. An introduction to quantum computing. Oxford University Press; 2007. 274 p.

75. Fedorov AK, Kiktenko EO, Lvovsky AI. Quantum computers put blockchain security at risk. Nature [Internet]. 2018 Nov 19 [cited 2019 Feb 16];563(7732):465–7. Available from: http://www.nature.com/articles/d41586-018-07449-z

76. Gheorghiu V, Gorbunov S, Mosca M, Munson B. Quantum-Proofing the Blockchain. Univ Waterloo [Internet]. 2017;(November). Available from: https://www.evolutionq.com/assets/mosca_quantum-proofing-the-blockchain_blockchain-research-institute.pdf

77. Stewart I, Ilie D, Zamyatin A, Werner S, Torshizi MF, Knottenbelt WJ. Committing to quantum resistance: a slow defence for Bitcoin against a fast quantum computing attack. R Soc Open Sci [Internet]. 2018 Jun 20;5(6):180410. Available from: https://royalsocietypublishing.org/doi/10.1098/rsos.180410

78. Parker L, Karliychuk T, Gillies D, Mintzes B, Raven M, Grundy Q. A health app developer's guide to law and policy: a multi-sector policy analysis. BMC Med Inform Decis Mak [Internet]. 2017 Dec 2 [cited 2019 Mar 8];17(1):141. Available from: http://www.ncbi.nlm.nih.gov/pubmed/28969704

79. Koczkodaj WW, Mazurek M, Strzałka D, Wolny-Dominiak A, Woodbury-Smith M. Electronic Health Record Breaches as Social Indicators. Soc Indic Res [Internet]. 2019;141(2):861–71. Available from: https://doi.org/10.1007/s11205-018-1837-z

80. HealthIT.gov. Guide to Privacy and Security of Health Information. 2013 [cited 2018 Feb 23];(April):27–40. Available from: https://www.healthit.gov/sites/default/files/pdf/privacy/privacy-and-security-guide.pdf

81. Jamshed N, Ozair FF, Sharma A, Aggarwal P, Jamshed N, Sharma A, et al. Ethical issues in electronic health records: A general overview. Perspect Clin Res [Internet]. 2015 [cited 2018 Feb 23];6(2):73. Available from: http://www.ncbi.nlm.nih.gov/pubmed/25878950

82. Liu Z, Luong NC, Wang W, Niyato D, Wang P, Liang Y, et al. A Survey on Applications of Game Theory in Blockchain. 2019 Feb;1–26.

83. Sonnino A, Gauba A, Apostolou D. A MetaAnalysis of Proposed Alternative Consensus Protocols for Blockchains. 2018;14(8):16. Available from: https://github.com/Mechanism-Labs/MetaAnalysis-of-Alternative-Consensus-Protocols

84. Attiya H, Welch J. Distributed Computing; Fundamentals, Simulations and Advanced Topics [Internet]. Second. Albert Y. Zomaya, editor. JOHN WILEY & SONS, INC; 2004. 432 p. Available from: https://www.wiley.com/en-us/Distributed+Computing%3A+Fundamentals%2C+Simulations%2C+and+Advanced+Topics%2C+2nd+Edition-p-9780471453246

85. Wang W, Hoang DT, Hu P, Xiong Z, Niyato D, Wang P, et al. A Survey on Consensus Mechanisms and Mining Strategy Management in Blockchain Networks. IEEE Access. 2019;7(c):22328–70.

86. Bano S, Sonnino A, Al-Bassam M, Azouvi S, McCorry P, Meiklejohn S, et al. Consensus in the Age of Blockchains. arXiv.org [Internet]. 2017 Nov 10; Available from: http://arxiv.org/abs/1711.03936

87. Schrepel T. Collusion By Blockchain And Smart Contracts. Harv J Law Technol [Internet]. 2019;33(1). Available from: https://www.ssrn.com/abstract=3315182

88. Mackey TK, Kuo TT, Gummadi B, Clauson KA, Church G, Grishin D, et al. "Fit-for-purpose?" - Challenges and opportunities for applications of blockchain technology in the future of healthcare. BMC Med. 2019;17(1):1–17.

89. Colón KA. Creating a Patient-Centered, Global, Decentralized Health System: Combining New Payment and Care Delivery Models with Telemedicine, AI, and Blockchain Technology. Blockchain Healthc Today. 2018;1:1–18.

90. Shen B, Guo J, Yang Y. MedChain: Efficient Healthcare Data Sharing via Blockchain. Appl Sci [Internet]. 2019;9(6):1207. Available from: https://www.mdpi.com/2076-3417/9/6/1207

91. Benchoufi M, Ravaud P. Blockchain technology for improving clinical research quality. Trials [Internet]. 2017 Dec 19;18(1):335. Available from: http://trialsjournal.biomedcentral.com/articles/10.1186/s13063-017-2035-z

92. Leeming G, Ainsworth J, Clifton DA. Blockchain in health care: hype, trust, and digital health. Lancet (London, England) [Internet]. 2019 Jun 22 [cited 2019 Jul 13];393(10190):2476–7. Available from: http://www.ncbi.nlm.nih.gov/pubmed/31232356